\documentclass[prl,superscriptaddress,showpacs,twocolumn]{revtex4}
\usepackage{graphicx}

\begin{document}

\title{Experimental demonstration of the time reversal Aharonov-Casher effect}
\affiliation{NTT Basic Research Labs, 3-1 Morinosato-Wakamiya, Atsugi-shi, Kanagawa 243-0198, Japan}
\affiliation{Graduate School of Engineering, Tohoku University, 6-6-02 Aramaki-Aza Aoba, Aoba-ku, Sendai 980-8579, Japan}
\affiliation{CREST-Japan Science and Technology Agency, Kawaguchi Center Building, 4-1-8, Hon-cho, Kawaguchi-shi, Saitama 332-0012, Japan}
\author{Tobias Bergsten}
\affiliation{NTT Basic Research Labs, 3-1 Morinosato-Wakamiya, Atsugi-shi, Kanagawa 243-0198, Japan}
\affiliation{CREST-Japan Science and Technology Agency, Kawaguchi Center Building, 4-1-8, Hon-cho, Kawaguchi-shi, Saitama 332-0012, Japan}
\author{Toshiyuki Kobayashi}
\author{Yoshiaki Sekine}
\affiliation{NTT Basic Research Labs, 3-1 Morinosato-Wakamiya, Atsugi-shi, Kanagawa 243-0198, Japan}
\author{Junsaku Nitta}
\affiliation{NTT Basic Research Labs, 3-1 Morinosato-Wakamiya, Atsugi-shi, Kanagawa 243-0198, Japan}
\affiliation{Graduate School of Engineering, Tohoku University, 6-6-02 Aramaki-Aza Aoba, Aoba-ku, Sendai 980-8579, Japan}
\affiliation{CREST-Japan Science and Technology Agency, Kawaguchi Center Building, 4-1-8, Hon-cho, Kawaguchi-shi, Saitama 332-0012, Japan}

\begin{abstract}
We demonstrate the time reversal Aharonov-Casher (AC) effect in small arrays of mesoscopic semiconductor rings. By using an electrostatic gate we can control the spin precession rate and follow the AC phase over several interference periods. We show that we control the precession rate in two different gate voltage ranges; in the lower range the gate voltage dependence is strong and linear and in the higher range the dependence in almost an order of magnitude weaker. We also see the second harmonic of the AC interference, oscillating with half the period. We finally map the AC phase to the spin-orbit interaction parameter $\alpha$ and find it is consistent with Shubnikov-de Haas analysis.
\end{abstract}

\pacs{85.35.Ds,73.23.-b,71.70.Ej}
\maketitle

Spintronics is the art of generating, manipulating and detecting the spin of electrons in solid state electronic devices. While this has traditionally involved ferromagnetic materials and external magnetic fields, we can also manipulate spins with purely electric fields via the spin-orbit interaction (SOI) between a moving spin particle and an electric field. In particular, we can design a semiconductor heterostructure with a two dimensional electron gas (2DEG) which has an internal electric field perpendicular to the 2DEG due to an asymmetric quantum well. We will then have SOI even without external electric fields. This is called the Rashba effect \cite{rashba,bychkov}.

The SOI is a relativistic effect on a particle with spin which is moving through an electric field. In the particle's frame of reference there will be a magnetic field perpendicular to the electric field and the direction of movement. The spin direction will precess around the axis parallel to this magnetic field and the precession rate depends on the spin-orbit interaction strength $\alpha$, and the value of $\alpha$ can be controlled by a gate voltage \cite{nittaSOI}. This allows us to control the spin precession rate with an electrostatic gate on top of the heterostructure.

In this letter we present evidence of quantum interference effects due to spin precession in small arrays of mesoscopic 2DEG rings. This interference is an Aharonov-Casher (AC) effect \cite{ahacash} of time reversal symmetric paths and is the electromagnetic dual \cite{mathur} to the Al'tshuler-Aronov-Spivak (AAS) effect \cite{aroaltspi}. As the AAS effect and the related Aharonov-Bohm (AB) effect have proven to be important tools in research, we can expect that the AC effect will be a powerful tool for understanding quantum interactions and material properties. We further show that we can control the spin precession rate with an electrostatic gate and modulate the interference pattern over several periods. Earlier experiments on square loop arrays yielded similar results, but only for up to one interference period \cite{kogaSpin}. We also see the second order AC effect where the oscillation period is half the period of the first order effect and the oscillations correspond to up to 26$\pi$ spin precession angle.
Second and third order harmonics have also recently been observed in square loop arrays \cite{sekine}. We map the interference pattern to changes in $\alpha$ and combine the results with measurements of $\alpha$ using Shubnikov-de Haas (SdH) beating patterns, which are consistent with the spin interference data. Other related experiments include single rings which gave inconclusive results \cite{nittaABAC}, and Aharonov-Bohm type AC effect in a single ring with a complex gate voltage dependence \cite{konig}.


\begin{figure}
\includegraphics[width=60mm]{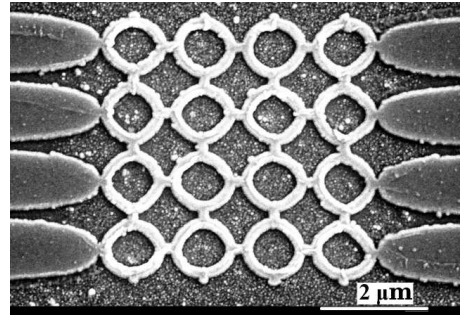}
\caption{An SEM image of an array of 0.5 $\mu$m radius rings. The array is covered by a 50 nm SiO$_2$ insulator layer and an 80 nm Au gate. The width of the rings is 100 nm after dry-etching.}
\label{SEMimage}
\end{figure}

The ring arrays were etched out in a dry-etching process from an InAlAs/InGaAs based 2DEG (Fig. \ref{SEMimage}), similar to sample 1 in Ref. \cite{kogaWAL}. The electron mobility was 7-11 m$^2$/Vs depending on the carrier density and the effective electron mass $m^*$ was 0.050$m_e$ as determined from the temperature dependence of SdH oscillations. By using arrays rather than single rings we get a stronger signal and we average out some of the universal conductance fluctuations (UCF) and AB oscillations \cite{umbach}. The arrays consisted of between 3$\times$3 and 6$\times$6 rings. The (average) radius of the rings was between 0.5 and 1.1 $\mu$m and the width was 20\% of the radius. The rings were covered with a 50 nm thick SiO$_2$ insulator layer, deposited by ECR sputtering, and on top of that was an Au gate, used to control the carrier density and the SOI strength $\alpha$. The advantage of using a small number of rings rather than a large array is that the gate leakage is much smaller and we can use relatively high gate voltage. This makes it possible to see several oscillations of AC interference.

Close to the arrays and in the same current path and under the same gate was a Hall bar, 5 $\mu$m wide and 20 $\mu$m long, used to measure the carrier density. For the SdH measurements we used a larger Hall bar, 20 $\mu$m by 120 $\mu$m, covered by an identical gate structure.


\begin{figure}
\includegraphics[width=86mm]{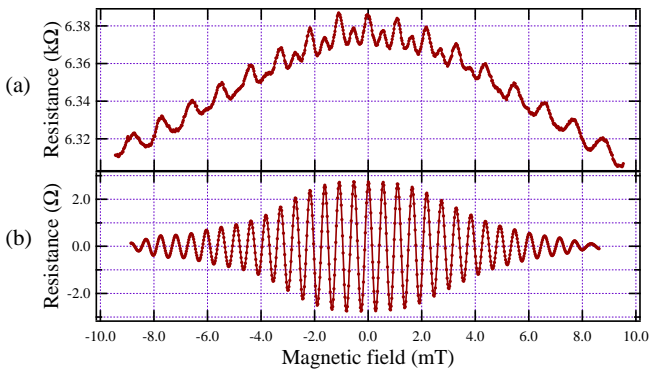}
\caption{The resistance of an array of rings versus the magnetic field. The array consisted of 4$\times$4 rings with 1.1 $\mu$m radius. (a) Raw data, averaged from eight sweeps with slightly different gate voltages (1.9-2.0 V).
(b) The same data after going through a digital band-pass filter, showing the AAS oscillations clearly with the amplitude falling off with higher field.}
\label{RvsB}
\end{figure}

The resistance of a mesoscopic ring is affected by various quantum interference effects. One is the AB effect which occurs when the electron wavefunction splits into two parts as they enter the ring and then interfere at the exit point. Depending on the magnetic flux $\Phi$ inside the ring the interference will be constructive or destructive and the resistance will oscillate with the period $2\Phi_0$ (here $\Phi_0=h/2e$ is the flux quantum). Note that the wavefunction phase is very sensitive to the scatterer configuration, so the interference is not necessarily constructive at zero magnetic flux.

Another effect is the AAS effect, which works in parallel with the AB effect and occurs when the two wavefunction parts travel a full turn around the ring in opposite directions and interfere at the entry point. Contrary to the AB effect, the two parts follow the exact same path but in different directions, or time reversal symmetric paths. Therefore the interference is always constructive in the absence of magnetic flux, meaning that the electron is scattered back and the resistance is increased. When the flux is increased the resistance oscillates with the period $\Phi_0$, but the amplitude decays after a few periods because of averaging between different paths in the ring, with different areas.


In Fig. \ref{RvsB} we display the resistance versus magnetic field for an array of rings. In the top graph we can see both interference effects described above. The fast oscillations near the center are the AAS oscillations and the slower are the AB oscillations which do not decay significantly at higher magnetic fields \cite{stone}. Then there is also \textit{weak localization} negative magnetoresistance \cite{bergmann} forming the wide peak on top of which the AB and AAS effects are superimposed. In our experiments we are interested in the AAS oscillations and in order to see them more clearly we feed the raw data through a digital band-pass filter, as shown in the lower graph. 

\begin{figure}
\includegraphics[width=86mm]{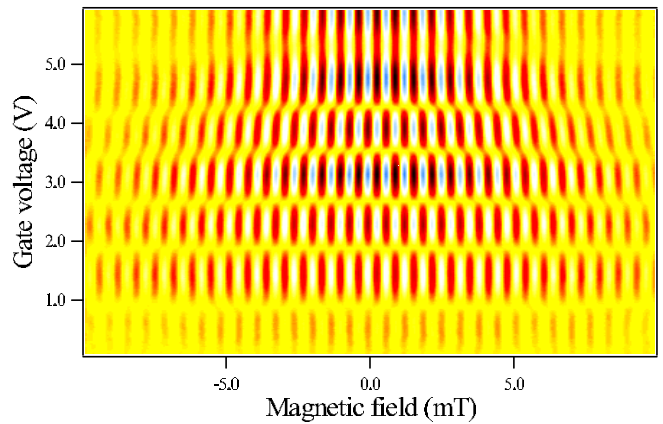}
\caption{The resistance vs gate voltage and magnetic field after digital filtering [see figure \ref{RvsB}(b)]. Colour scale is black-red-yellow-white-blue. We clearly see the AAS oscillations switching phase as the gate voltage is increased. The sample is a 5$\times$5 array of 1.0 $\mu$m radius rings.}
\label{RvsBVgGray}
\end{figure}

If there is SOI in the ring, the electron spin will start precessing around the effective magnetic field and change the interference at the entry point. The precession axes for the two parts of the wavefunction are opposite and therefore the relative precession angle (the AC phase) is twice the angle of each part. If the relative precession angle is $\pi$ the spins of the two parts are opposite and can not interfere, and the AAS oscillations disappear. If the relative angle is $2\pi$ the two parts will have the same spin but opposite signs because of the 1/2 spin quantum laws (a 4$\pi$ rotation is required to return to the original wavefunction), effectively changing the phase of the AAS oscillations by $\pi$, which we interpret as a negative amplitude.

In the following, precession angle (rate) refers to the precession of a single path.

The precession angle of an electron moving along a straight narrow channel is \cite{DattaDas}
\begin{equation}\label{Theta}
	\theta_0=\frac{2\alpha m^*}{\hbar^2}L,
\end{equation}
with $m^*$ being the effective electron mass and $L$ the distance travelled. In a ring the precession is a bit more complicated because of the precession axis constantly changing direction. The AAS interference amplitude can be written as \cite{nittaSID,Frustaglia,Veenhuizen}
\begin{equation}\label{AAS}
	\frac{\delta R_{\alpha}}{\delta R_{\alpha=0}}=
	\cos\left(2\pi\sqrt{1+\left(\frac{2\alpha m^*}{\hbar^2}r\right)^2}\right),
\end{equation}
with $\delta R_{\alpha}$ and $\delta R_{\alpha=0}$ being the AAS amplitude with and without SOI, respectively, and $r$ the radius of the ring. In the limit of strong SOI or large rings the argument of the cosine reduces to $\theta_0$ because the distance travelled around the ring is $2\pi r$.
%
%


\begin{figure}
\includegraphics[width=86mm]{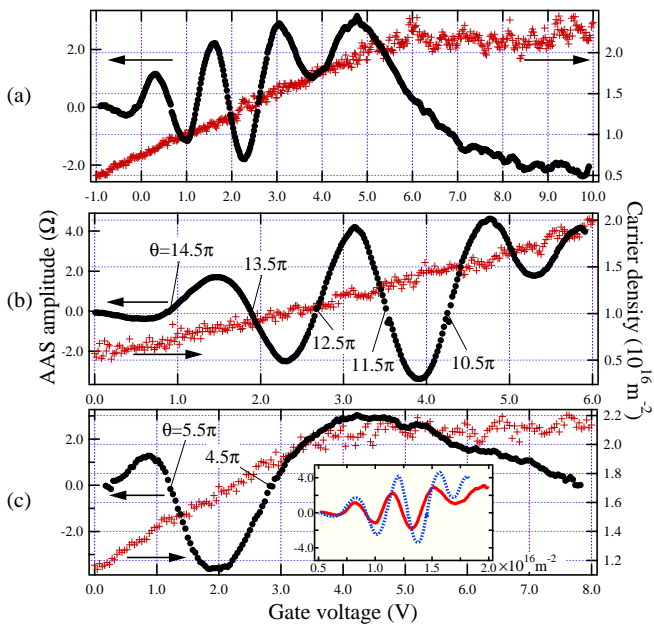}
\caption{The time reversal AC effect. AAS amplitude and carrier density plotted against the gate voltage for three different ring arrays: (a) 3$\times$3 array, 1.0 $\mu$m radius; (b) 5$\times$5 array, 1.0 $\mu$m; (c) 6$\times$6 array, 0.5 $\mu$m. Inset of (c): the amplitudes of (a) [solid] and (b) [dotted] plotted against the carrier density. The precession angle $\theta$ corresponds to the argument of the cosine in Eq. (\ref{AAS}).}
\label{AAS_NevsV}
\end{figure}

The experiment was carried out in a $^3$He cryostat at the base temperature which varied between 220 and 270 mK. The sample was in the core of a superconducting magnet with the field $B$ perpendicular to the 2DEG plane. We measured the resistance $R$ of the ring array simultaneously with the Hall resistance $R_H$ of the Hall bar close to the rings, while stepping the magnetic field and the gate voltage $V_G$. Figure \ref{RvsBVgGray} shows the result after digital band-pass filtering of the AAS oscillations which are visible as vertical bands in the figure. We can clearly see the oscillations switching phase as we increase the gate voltage.

In order to reduce noise and UCF effects we averaged ten resistance versus magnetic field (R vs B) curves with slightly different gate voltages. This averaging preserves the AAS oscillations but the averaging of $M$ curves reduces the AB amplitude roughly as $M^{-1/2}$ \cite{nittaABAC,morpurgo}.



We calculated the AAS amplitude by integration of the FFT spectra of $R$ vs $B$ curves. We then plotted the amplitude against the gate voltage. Figure \ref{AAS_NevsV} shows the results from three different ring arrays. We also calculated the carrier density $n_e$ from the slope of the $R_H$ vs $B$ ($n_e^{-1}=e\ dR_H/dB$) and plotted it in the same diagrams. It was important to measure $n_e$ simultaneously with $R$ rather than measuring the $n_e$ vs $V_G$ dependence separately because the dependence shifted considerably between different gate voltage sweeps. This is obvious from Fig. \ref{AAS_NevsV}(a) and (b) which are measured on arrays with the same ring radius, 1.0 $\mu$m. All the features of graph (b) are faithfully reproduced in graph (a), except it is shifted by about $\Delta V_G=$1 V. If we plot the amplitude against the carrier density instead [inset of Fig. \ref{AAS_NevsV}(c)] we see that the two curves agree very well.

As we see in Fig. \ref{AAS_NevsV} the AAS amplitude oscillates as we change the carrier concentration using the top gate. As the gate voltage is changed, the SOI strength $\alpha$ changes with it and as we expect from Eq. (\ref{AAS}) the AAS amplitude crosses zero, inverting the AAS oscillations. Each period represents one extra 2$\pi$ spin precession of an electron moving around a ring. However, the top two graphs has an unexpected `half oscillation' (a negative peak which does not cross zero) at a carrier concentration of 1.7$\times10^{16}$ m$^{-2}$. There is no theoretical explanation for this result. We also notice from the top and bottom graphs that the carrier density saturates around 2.2$\times10^{16}$ m$^{-2}$, and that the AAS amplitude is still changing, but much more slowly.

\begin{figure}
\includegraphics[width=86mm]{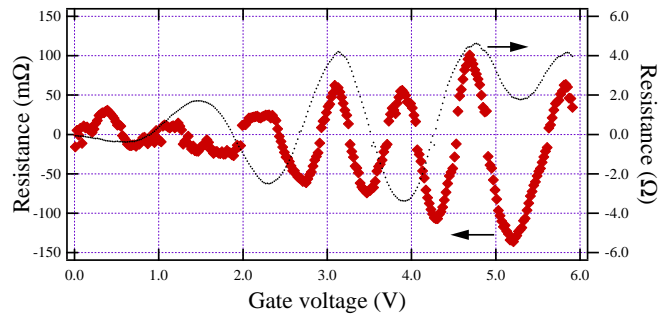}
\caption{The second harmonic AAS oscillation amplitude (diamonds) oscillates with half the period compared to the first harmonic AAS effect (dots). This data is from the same measurement as Fig. \ref{AAS_NevsV}(b).}
\label{AAS2ndharm}
\end{figure}

In the frequency spectra there is also a small peak at twice the AAS frequency. This is due to the wavefunction parts going twice around the ring before interfering. If we do the same analysis on this peak we get an oscillating amplitude with half the period compared to the first harmonic AAS amplitude (Fig. \ref{AAS2ndharm}). This is expected because the distance is twice and therefore the precession angle is also twice.
The first peak we can distinguish from the noise, at 2.3 V, corresponds to a precession angle of 26$\pi$.
In the second harmonic curve the negative peaks correspond to the zero crossings of the first harmonic, while the positive peaks correspond to the peaks, positive and negative. We note one interesting point: the negative peak at 5.2 V in the second harmonic corresponds to the `half oscillation' in the first harmonic rather than a zero crossing.

We can use the oscillations to map $\alpha$ against $n_e$, like in ref. \onlinecite{kogaSpin}. According to Eq. (\ref{AAS}) the AAS amplitude is zero when
\begin{equation}\label{alphazero}
	\alpha=\pm\frac{\hbar^2}{8m^*r}\sqrt{\left(2N+1\right)^2-16},
\end{equation}
with integer $N>1$. The zero crossings in the graphs correspond to consecutive values of $N$. However, this does not tell us the absolute value of $\alpha$. We need some way to anchor the string of zero crossings to an absolute value. One established way of doing this is to measure SdH oscillations in the sheet magnetoresistance \cite{schapersSdH}. 
%



\begin{figure}
\includegraphics[width=86mm]{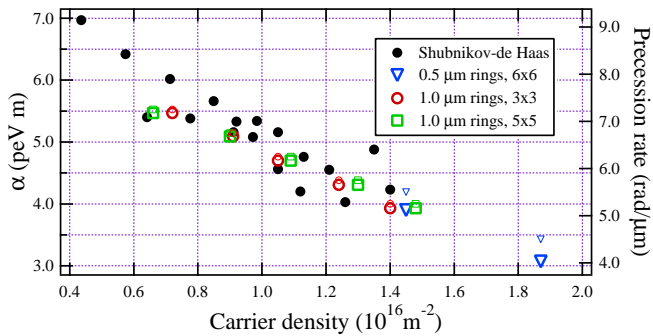}
\caption{By combining $\alpha$ values derived from Shubnikov-de Haas measurements (dots) and from the zero crossings of the spin interference graphs (squares, rings and triangles) we can map $\alpha$ values over a wide range of carrier densities. The right axis shows the corresponding precession rates according to Eq. (\ref{Theta}). The small symbols are the actual precession rates in the rings (the argument of the cosine in Eq. (\ref{AAS})) which are slightly higher, particularly for the 0.5 $\mu$m rings.}
\label{AlphavsNe}
\end{figure}

We measured SdH oscillations in a wide range of carrier densities in a separate Hall bar. The beating pattern was not very pronounced but we could get rough values of $\alpha$ (black dots in Fig. \ref{AlphavsNe}). We know from similar heterostructures that the value of $\alpha$ is positive \cite{kogaWAL} so we can drop the sign uncertainty from Eq. (\ref{alphazero}). Now we can map the spin interference graphs to absolute $\alpha$ values. There is one common zero crossing in all three graphs, close to 1.45$\times10^{16}$ m$^{-2}$. We can anchor this point to the SdH values by choosing $N=10$ and $N=5$ for the 1.0 $\mu$m and 0.5 $\mu$m radius arrays respectively, see Fig. \ref{AlphavsNe}. The gate voltage sensitivity $\Delta\alpha/\Delta V_G$ is 0.46-0.57 peV\,m/V in the range below carrier density saturation. In the saturation region the sensitivity is much smaller. In Fig. \ref{AAS_NevsV}(a) and \ref{AAS_NevsV}(c) the precession angle changes by roughly $\pi/2$ over 4 V, which gives a sensitivity of 0.05 and 0.11 peV\,m/V respectively.

To conclude, we have shown that the spin precession rate can be controlled in a precise and predictable way with an electrostatic gate. We have experimentally demonstrated the time reversal AC effect, the electromagnetic dual to the AAS effect, in small arrays of rings, including the second harmonic of this effect. We have also shown that as we saturate the carrier density in the rings, we can still control the precession rate, but with a much lower sensitivity. Our spin interference data agree with SdH measurements of $\alpha$. The precise spin precession control is important in order to realize semiconductor spintronics devices based on SOI.

\end{document}